\documentclass[fleqn,usenatbib]{mnras}

\usepackage{newtxtext,newtxmath}

\usepackage[T1]{fontenc}

\DeclareRobustCommand{\VAN}[3]{#2}
\let\VANthebibliography\thebibliography
\def\thebibliography{\DeclareRobustCommand{\VAN}[3]{##3}\VANthebibliography}

\usepackage{graphicx}	
\usepackage{amsmath}	

\usepackage[dvipsnames]{xcolor}

\title[Multiplicity in the $\dot{M}_\mathrm{acc}-M_\mathrm{disc}$ plane]{Stellar multiplicity affects the correlation between proto-planetary disc masses and accretion rates: binaries explain high-accretors in Upper Sco}

\author[F.~Zagaria et al.]{
Francesco Zagaria,$^{1}$\thanks{E-mail: fz258@cam.ac.uk}
Cathie J. Clarke,$^{1}$
Giovanni P. Rosotti,$^{2,3}$
and Carlo F. Manara$^{4}$ 
\\
$^{1}$Institute of Astronomy, University of Cambridge, Madingley Road, Cambridge CB3 0HA, UK\\
$^{2}$School of Physics and Astronomy, University of Leicester, Leicester LE1 7RH, UK\\
$^{3}$Leiden Observatory, Leiden University, PO Box 9513, 2300 RA Leiden, The Netherlands\\
$^{4}$European Southern Observatory, Karl-Schwarzschild-Strasse 2,
85748 Garching bei München, Germany\\
}

\date{Accepted XXX. Received YYY; in original form ZZZ}

\pubyear{2021}

\begin{document}
\label{firstpage}
\pagerange{\pageref{firstpage}--\pageref{lastpage}}
\maketitle

\begin{abstract}
In recent years a correlation between mass accretion rates onto new-born stars and their proto-planetary disc masses was detected in nearby young star-forming regions. Although such a correlation can be interpreted as due to viscous-diffusion processes in the disc, highly-accreting sources with low disc masses in more evolved regions remain puzzling. In this paper, we hypothesise that the presence of a stellar companion truncating the disc can explain these outliers. Firstly, we searched the literature for information on stellar multiplicity in Lupus, Chamaeleon~I and Upper Sco, finding that roughly 20 per cent of the discs involved in the correlation are in binaries or higher-order multiple stellar systems. We prove with high statistical significance that at any disc mass these sources have systematically higher accretion rates than those in single-stars, with the bulk of the binary population being clustered around $M_\mathrm{disc}/\dot{M}_\mathrm{acc}\approx0.1\,\mathrm{Myr}$. We then run coupled gas and dust one-dimensional evolutionary models of tidally truncated discs to be compared with the data. We find that these models are able to reproduce well most of the population of observed discs in Lupus and Upper Sco, even though the unknown eccentricity of each binary prevents an object by object comparison. In the latter region, the agreement improves if the grain coagulation efficiency is reduced, as may be expected in discs around close binaries. Finally, we mention that thermal winds and sub-structures can be important in explaining few outlying sources. 
\end{abstract}



\section{Introduction}\label{sec:1}
Understanding how proto-planetary discs evolve and eventually dissipate is fundamental to building a comprehensive picture of the planet-formation process \citep[e.g.,][]{Manara2019_pps}. It is commonly assumed that proto-planetary discs evolve under the effect of viscosity, which, allowing for angular momentum re-distribution within the disc, drives accretion onto the forming star \citep{Shakura&Sunyaev1973,Lynden-Bell&Pringle1974}. In this framework, the mass accretion rate on the central star, $\dot{M}_\mathrm{acc}$, and the disc mass, $M_\mathrm{disc}$, are expected to be correlated at any given age \citep[e.g.,][]{Hartmann1998,Jones2012}:
\begin{equation}\label{eq:1}
    M_\mathrm{disc}=t_\mathrm{acc}\dot{M}_\mathrm{acc}\propto(t+t_\nu)\dot{M}_\mathrm{acc},
\end{equation}
where $t_\nu$ is the initial viscous time scale and we call $t_\mathrm{acc}$ the accretion time scale\footnote{This is also known as the disc lifetime \citep[e.g.,][]{Lodato2017}.}.

The advent of new facilities, such as the Atacama Large Millimeter/submillimeter Array (ALMA) and the X-shooter spectrograph at the Very Large Telescope (VLT), has made it possible to test disc evolutionary models observationally. In the last years, disc masses, estimated from (sub-)millimetre dust fluxes under the assumption of optically thin emission and a constant gas-to-dust ratio \citep[e.g.,][]{Ansdell2016,Ansdell2018,Pascucci2016,Barenfeld2016}, as well as mass accretion rates \citep[e.g.,][]{Alcala2014,Alcala2017,Manara2016_cham,Manara2017,Manara2020}, have been inferred for a large number of young stellar objects. 

Combining these data-sets, \citet{Manara2016} and \citet{Mulders2017} detected a slightly sub-linear correlation between mass accretion rates and proto-planetary disc masses in the young Lupus and Chamaeleon~I star-forming regions (aged $\approx1\text{ to }3\,\mathrm{Myr}$), that was interpreted as being compatible with viscous-diffusion models \citep{Rosotti2017,Lodato2017,Mulders2017}. A correlation was also identified by \citet{Manara2020} in the older Upper Scorpius OB association (aged $\approx5\text{ to }10\,\mathrm{Myr}$). However, at any given disc mass, the median accretion rate in Upper Sco is remarkably similar to that in young regions; moreover its scatter has not decreased with time. This is in contrast with the viscous evolution scenario, which predicts both lower $\dot{M}_\mathrm{acc}$ and a tighter relation at later times (\citealt{Rosotti2017,Lodato2017} and Eq.~\ref{eq:1}). 

Nevertheless, all the previous analyses of the correlation only focused on \textit{gas} evolution. Instead, the disc mass estimates of \citet{Manara2016,Manara2020} and \citet{Mulders2017} are based on (sub-)millimetre \textit{dust} emission and are subject to systematic uncertainties (e.g., on dust opacity and the gas-to-dust ratio). 
To try to explain the observations, both gas and dust ought to be considered when modelling disc evolution. Moreover, theoretical predictions and observational trends must be compared in the \textit{data space}, i.e. post-processing the model outputs to determine their (sub-)millimetre dust fluxes and masses as for real data. Such an exercise then removes the use of the restrictive assumptions about optical depth, opacities and the gas-to-dust ratio that are generally used to interpret (sub-)millimetre observations. \citet{Sellek2020} adopted such a forward modelling approach and demonstrated that viscous simulations provide a remarkable agreement with both Lupus and Upper Sco data (in particular when internal photo-evaporation is included to account for late-time disc dispersal). However, even though the models of \citet{Sellek2020} accounted for the underestimation of disc masses in current surveys, none of them were able to reproduce discs with accretion time scale shorter than $t_\mathrm{acc}\lesssim0.1\,\mathrm{Myr}$, which contribute to most of the large scatter/high accretion rates in Upper Sco.

As mass accretion rates are computed at a fixed snapshot in time, it might be thought that accretion variability is responsible for the outlying sources. Unfortunately, there are no studies addressing this issue at the age of Upper Sco \citep{Manara2020}, and very little is known about accretion variability on time scales longer than few days. In younger ($t\lesssim3\,{\rm to}\,4\,{\rm Myr}$) discs, $\dot{M}_{\rm acc}$ variations are $\lesssim0.4\,{\rm dex}$ on a time scale of days \citep{Costigan2014,Venuti2014,Manara2021}, not enough to account for the highest accretors in Upper Sco, with only a small fraction of the targets having larger variability \citep{Audard2014}.

All the previous considerations are based on the assumption that planet-forming discs evolve in isolation. This picture is clearly idealised: the majority of stars are born in clusters, where nearby sources can influence proto-planetary disc evolution history and affect their planet-formation potential \citep[e.g.,][]{Winter2020}. This commonly occurs by either external photo-evaporation, when the UV radiation of a massive nearby star dissipates the less bound gas in the outer disc regions \citep[e.g.,][]{Adams2004,Facchini2016}, or by tidal interactions, when a stellar companion in a gravitationally bound pair, or a flyby in a dense environment, truncates the disc \citep[e.g.,][]{Winter2018,Cuello2019,Cuello2020}. These processes have qualitatively similar effects, both reducing disc masses and sizes (e.g., \citealt{Harris2012,Akeson&Jensen2014,Cox2017,Akeson2019,Manara2019} in binaries and \citealt{Ansdell2017,Otter2021} for the effects of photo-evaporation), eventually hastening disc dispersal.

How external photo-evaporation and tidal truncation impact the correlation between mass accretion rates and disc masses was studied by \citet{Rosotti2017} and \citet{Rosotti&Clarke2018}, respectively. These studies showed that both discs exposed to strong UV fields and discs in multiple stellar systems are expected to have shorter accretion time scales than those (viscously) evolving in isolation. This means that the influence of the environment could be a potential explanation for those highly-accreting old discs with low masses that the models of \citet{Sellek2020} were not able to explain. However, these studies only focused on gas evolution. Since dust was not considered as a separate component, it was simply assumed that observationally inferred disc dust masses could be converted into total disc masses using a gas-to-dust ratio of 100. This motivates our present study based on modelling both gas and dust evolution and then comparing simulation-based synthetic observations with real data.

The required properties of star-forming environments for which  external photo-evaporation and tidal encounters are effective were studied by \citet{Winter2018_trunc}. They showed that stellar densities $\gtrsim10^4\,{\rm pc}^{-3}$ are necessary for disc truncation by tidal encounters. This threshold is much higher than in the nearby star-forming regions, making fly-bys unlikely. As for external photo-evaporation, \citet{Winter2018_trunc} concluded that average UV fields $\gtrsim3\times10^3\,G_0$ are required to disperse a disc. (Here $G_0$ corresponds to an energy flux of $1.6\times10^{-3}\,{\rm erg}\,{\rm cm}^{-2}$ between $6\,{\rm and}\,13.6\,{\rm eV}$, \citealt{Habing1968}.) In Lupus the average radiation field is expected to be remarkably low (e.g., in IM~Lup it is $\lesssim4\,G_0$, \citealt{Cleeves2016}) while in Upper Sco it is somewhat higher (on average $42.9\,G_0$, \citealt{Trapman2020}). Since the median UV field in both regions is below the critical value for evaporation, we decided not to consider external photo-evaporation in this paper\footnote{In Upper Sco the radiation levels can grow up to $7\times10^3\,G_0$ \citep{Trapman2020}, suggesting that external photo-evaporation can influence the accretion time scales of some sources. We plan to study this effect in a future paper.}.

Instead, in this work we focus on the role played by stellar multiplicity in disc evolution, confronting our theoretical expectations with the mass accretion rates and sub-millimetre fluxes from nearby young star-forming regions. First of all, from the currently published catalogues, we collect a state-of-the-art sample of homogeneously determined disc masses and accretion rates in Lupus, Chamaeleon~I and Upper Sco. This sample is further updated making use of Gaia EDR3 distances \citep{GAIAeDR3} as in the PPVII chapter of Manara et al. (subm.). Furthermore, searching the literature allows us to identify several pairs, i.e. gravitationally bound stars with projected separation, $a_\mathrm{p}$, less than $300\,\mathrm{au}$, in binaries or higher order multiple stellar systems among the previously selected sources. Subsequently, we run models of tidally truncated circumstellar binary discs for a large number of initial disc parameters following the approach of \citet{Zagaria2021_theo}. To test our hypothesis that the accretion time scale is influenced by a companion, these are then compared with observations through modelling the predicted dust emission.

This paper is organised as follows. In Section~\ref{sec:2} we introduce the Lupus, Chamaeleon~I and Upper Sco samples. Discs in multiple stellar systems are identified (see Appendix~\ref{app:1}) and their properties discussed in comparison to those in single-star systems. In Section~\ref{sec:3} dust and gas modelling is introduced, while Section~\ref{sec:4} describes our results,  first of all discussing their dependence on  parameter choices and then comparing  models and observations. In Section~\ref{sec:5} we consider the model limitations and possible improvements to the agreement with observational data. Finally, in Section~\ref{sec:6} we draw our conclusions.

\section{Sample selection and data analysis}\label{sec:2}
\subsection{Sample selection} 
We focus on Lupus, Chamaeleon~I and Upper Sco because they are among the best studied nearby star-forming regions: (sub-)millimetre dust fluxes and accretion rates were measured for a large fraction of their young stellar objects (YSOs), and their multiplicity fraction is often well known.

The distances to individual targets are taken from inverting the parallaxes available in the Gaia EDR3 catalog \citep{GAIAeDR3}. The only exception are the cases where the Gaia parallaxes are unreliable, i.e. ${\rm RUWE}>1.8$ and/or distance differing more than $60\,{\rm pc}$ from the median distance to the region, or unavailable. In these cases, we assumed the median distance to the members of the region: $d=158\,{\rm pc}$ for Lupus, $d=190\,{\rm pc}$ for Chamaeleon~I, and $d=145\,{\rm pc}$ for Upper Sco.

In Lupus, we make use of the (sub-)millimetre dust emission measured by \citet{Ansdell2016}, with the addition of the Lupus completion survey sources of \citet{Sanchis2020}, Sz~82/IM~Lup \citep{Cleeves2016} and Sz~91 \citep{Tsukagoshi2019}. 
We convert the ALMA (sub-)millimetre dust fluxes into dust masses assuming optically thin emission \citep{Hildebrand1983}:
\begin{equation}\label{eq:2}
    M_\mathrm{dust}=\dfrac{d^2F_\nu}{B_\nu(T_\mathrm{dust})\kappa_\nu^\mathrm{abs}},\text{ where }\kappa_\nu^\mathrm{abs}=2.3\left(\dfrac{\nu}{230~\mathrm{GHz}}\right)\text{cm}^2\text{g}^{-1}.
\end{equation}
Here $F_\nu$ is the dust flux, $d$ the distance of the source, $B_\nu$ the black body emission at temperature $T_\mathrm{dust}=20\,\mathrm{K}$, and $\kappa_\nu^\mathrm{abs}$ the absorption opacity \citep{Beckwith1990}. Disc masses are then computed from dust masses assuming a standard ISM gas-to-dust ratio of 100 \citep{Bohlin1978}. A target is a non-detection if its dust flux does not exceed three times the continuum rms noise. 

Mass accretion rates are inferred from the VLT/X-shooter spectroscopic measurements of \citet{Alcala2014,Alcala2017}. Briefly, stellar properties and the accretion parameters are derived self-consistently fitting the YSO spectra with the sum of the photospheric template of a non-accreting star and a slab model to reproduce the accretion luminosity excess in the Balmer continuum and also in the optical part of the spectrum, as described in \citet{Manara2013}. Following Manara et al. (subm.), the effective temperature, $T_{\rm eff}$, of the stars were re-calculated using the relation between spectral type and $T_{\rm eff}$ by \citet{Herczeg&Hillenbrand2014}, and the stellar luminosity and accretion luminosity were re-scaled to the distances from Gaia EDR3. To derive the stellar masses, needed to calculate the mass accretion rates, the non-magnetic evolutionary tracks by \citet{Feiden2016} and \citet{Baraffe2015} where used for targets hotter and colder than 3900 K, respectively, while the models by \citet{Siess2000} were used for sources appearing too young/old for the former tracks. Under-luminous objects (see \citealt{Alcala2014}), i.e., falling well below the main locus of pre-main sequence targets in the HR diagram, are excluded. If the excess continuum UV luminosity of a YSO with respect to the best-fit photospheric template is close to its chromospheric noise level, it is termed a non-accretor and its mass accretion rate is considered an upper-limit \citep[see][]{Alcala2017,Manara2017}. The accretion rate of MY~Lup is as computed from the HST observation results of \citet{Alcala2019}. After updating stellar distances, an average uncertainty of $0.35\,{\rm dex}$ is estimated.

In Chamaeleon~I and Upper Sco we follow a similar procedure. In the former region, we use the (sub-)millimetre dust emission of \citet{Pascucci2016} and \citet{Long2018_cham}, and the accretion luminosities of \citet{Manara2016_cham,Manara2017}. In the latter region, we rely on \citet{Manara2020} sample for mass accretion rates, while disc masses are computed using the (sub-)millimetre dust emission of \citet{Barenfeld2016}. 

Transition discs (TDs) are identified as both discs with a ``resolved'' (imaged) cavity and ``unresolved'' discs, whose classification is based on their SED shape as in \citet{Manara2016,Manara2020} and \citet{Mulders2017}. Finally, we remark that Lupus and Chamaeleon~I samples are more than 90 per cent complete\footnote{In this paper, by saying that a given star-forming region is $x$ per cent complete, we mean that disc accretion rates were measured for a fraction $x/100$ of the YSOs with discs in that region.} \citep{Alcala2017,Manara2017}, while Upper Sco has high levels of completeness only for $M_\mathrm{disc}\gtrsim5\times10^{-5}\,M_\mathrm{Sun}$ \citep{Manara2020}\footnote{More specifically, \citet{Manara2020} claimed completeness to be 80 per cent in the mass range $4.8\times10^{-5}\,M_\mathrm{Sun}\leq M_\mathrm{disc}\leq6.47\times10^{-4}\,M_\mathrm{Sun}$ and 60 per cent in the mass range $6.47\times10^{-4}\,M_\mathrm{Sun}\leq M_\mathrm{disc}\leq1.55\times10^{-2}\,M_\mathrm{Sun}$ \citep{Manara2020}. Nevertheless, these are likely overestimates if the refined membership census of YSOs with discs in Upper Sco \citep{Luhman&Esplin2020} is considered.}.

We searched the literature for information on stellar multiplicity for the previously selected YSOs. Our main references are \citet{Zurlo2021} in Lupus, \citet{Lafreniere2008} in Chamaeleon~I and \citet{Barenfeld2019} in Upper Sco. A total of 9 binaries or higher-order multiple stellar systems with projected separation less than $300\,\mathrm{au}$ are found is Lupus, 16 in Chamaeleon and 9 in Upper Sco. 11 of these are in pairs closer than $40\,\mathrm{au}$ \citep{Kraus2012}. We refer the reader to Appendix~\ref{app:1} for a detailed discussion of the multiplicity detection methods and comparison among regions.

\subsection{Data analysis}

\begin{figure*}
    \centering
    \includegraphics[width=\textwidth]{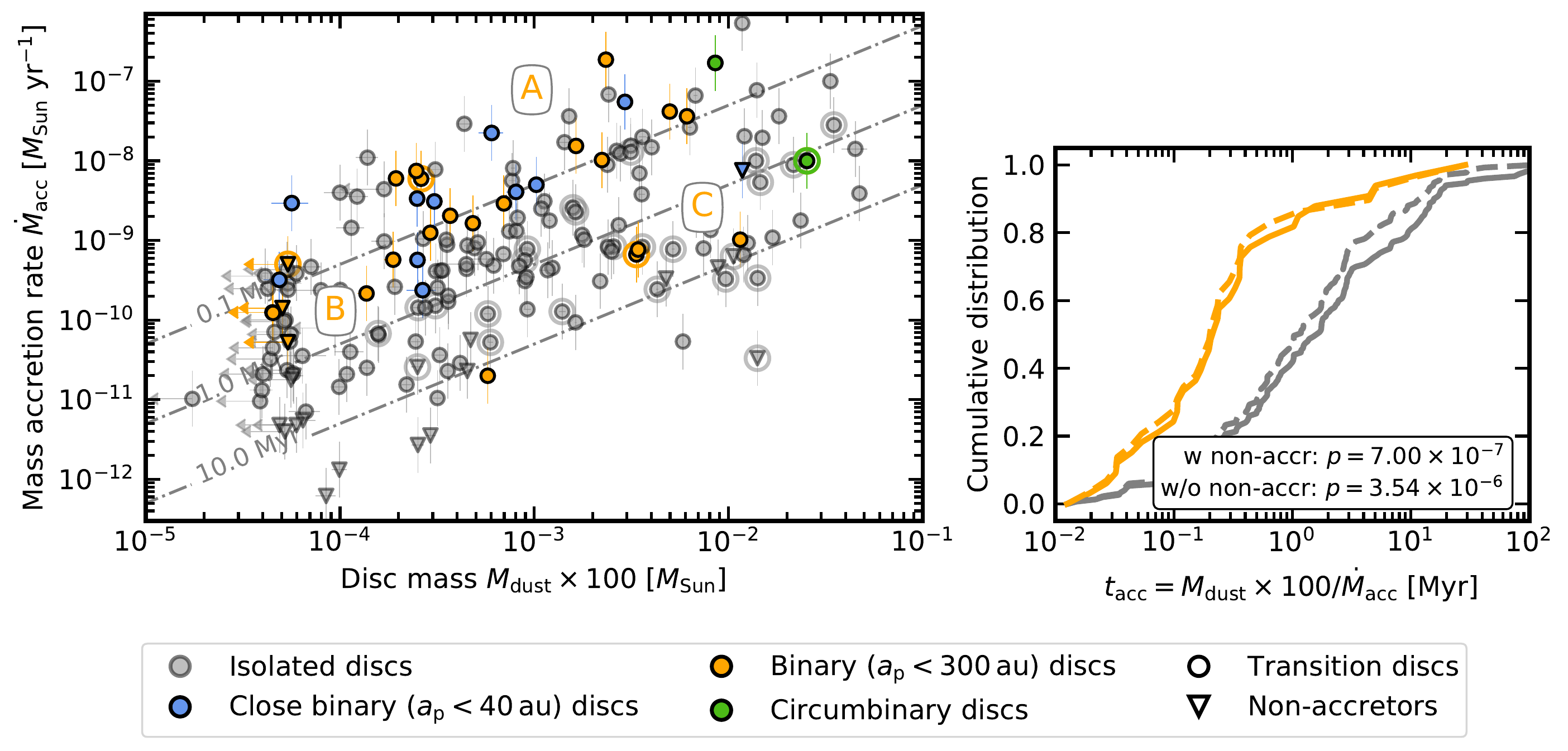}
    \caption{\textbf{Left panel:} Mass accretion rate, $\dot{M}_\mathrm{acc}$, as a function of the disc mass, $M_\mathrm{disc}=M_\mathrm{dust}\times100$, for Lupus, Chamaeleon~I and Upper Sco Class~II sources. Grey dots are used for isolated stars, orange and blue for binary ($a_\mathrm{p}<300\,\mathrm{au}$) and close binary ($a_\mathrm{p}<40\,\mathrm{au}$) discs, and green for circumbinary discs. Transition discs are identified by a wider circle, non-accretors by a downward-pointing triangle and non-detections by a left-pointing arrow. In the background, lines of constant $t_\mathrm{acc}=0.1,\,1$ and $10\,\mathrm{Myr}$ are shown as dot-dashed lines. Letters A, B and C are used to label different binary disc populations (see main text). \textbf{Right panel:} Accretion time scale, $t_\mathrm{acc}$, cumulative distribution for single-star and binary discs in grey and orange, respectively. Solid lines are used for the full sample, dashed lines when non-accretors are excluded. The caption lists $p$-values for the null hypothesis of binaries and singles being drawn from the same $t_{\rm acc}$ distribution.}
    \label{fig:2.1}
\end{figure*}

\paragraph*{General population trends} In the left-hand panel of Fig.~\ref{fig:2.1} the mass accretion rate, $\dot{M}_\mathrm{acc}$, is plotted as a function of the disc mass inferred from (sub-)millimetre dust emission, $M_\mathrm{disc}=M_\mathrm{dust}\times100$, for all Lupus, Chamaeleon and Upper Sco discs in our sample. Grey dots are used for isolated stars, orange and blue for binary ($a_\mathrm{p}<300\,\mathrm{au}$) and close binary ($a_\mathrm{p}<40\,\mathrm{au}$) discs, while green for circumbinary discs. Non-accretors are displayed as downward-pointing triangles, while disc mass non-detections are identified by left-pointing arrows. Adopting the same convention of \citet{Manara2016}, transition discs can be recognised by a larger external circle of the same colour. In the background, lines of constant accretion time scale for $t_\mathrm{acc}=0.1,\,1\,\mathrm{and}\,10\,\mathrm{Myr}$ are displayed as grey dashed-dotted lines.

Single-star discs and discs in binary systems populate different regions of the data-space. The former span all the range of observed disc masses and accretion rates, while the latter are characterised by a systematically shorter accretion time scale and can be divided into three sub-regions. The bulk of discs in binaries (A) have $t_\mathrm{acc}\approx0.1\,\mathrm{Myr}$. The remainder have longer accretion time scales, with $t_\mathrm{acc}\gtrsim1\,\mathrm{Myr}$. Some of these are very faint (B), often undetected in the (sub-)millimetre, with $M_\mathrm{disc}\lesssim2\times10^{-4}\,M_\mathrm{Sun}$, while a small number are massive (C), with $M_\mathrm{disc}\gtrsim2\times10^{-3}\,M_\mathrm{Sun}$. Remarkably, very close binaries ($a_\mathrm{p}<40\,\mathrm{au}$) mainly live in region A, and $5/11\approx45$ per cent have accretion time scales much shorter than $0.1\,\mathrm{Myr}$, because of high accretion rates and (generally) low 
disc masses.

To prove our inference of a shorter accretion time scale in discs in multiple systems we perform a Kolmogorov–Smirnov (K-S) test on the full sample made up of all data from the three star-forming regions together, excluding the three circumbinary discs in the sample. The null-hypothesis of single-star and binary discs being drawn from the same $t_\mathrm{acc}$ distribution is rejected with $p-$values of $7.00\times10^{-7}$ and $3.54\times10^{-6}$, when non-accretors are considered or excluded, respectively. This is visually displayed in the right-hand panel of Fig.~\ref{fig:2.1}, where the accretion time scale cumulative distributions are plotted for single-star and binary discs in grey and orange, respectively. Solid lines are used for the full sample, while dashed ones when non-accretors are not considered. Clearly, discs in multiple systems have systematically shorter $t_\mathrm{acc}$ than those evolving in isolation.

\begin{figure*}
    \centering
    \includegraphics[width=\textwidth]{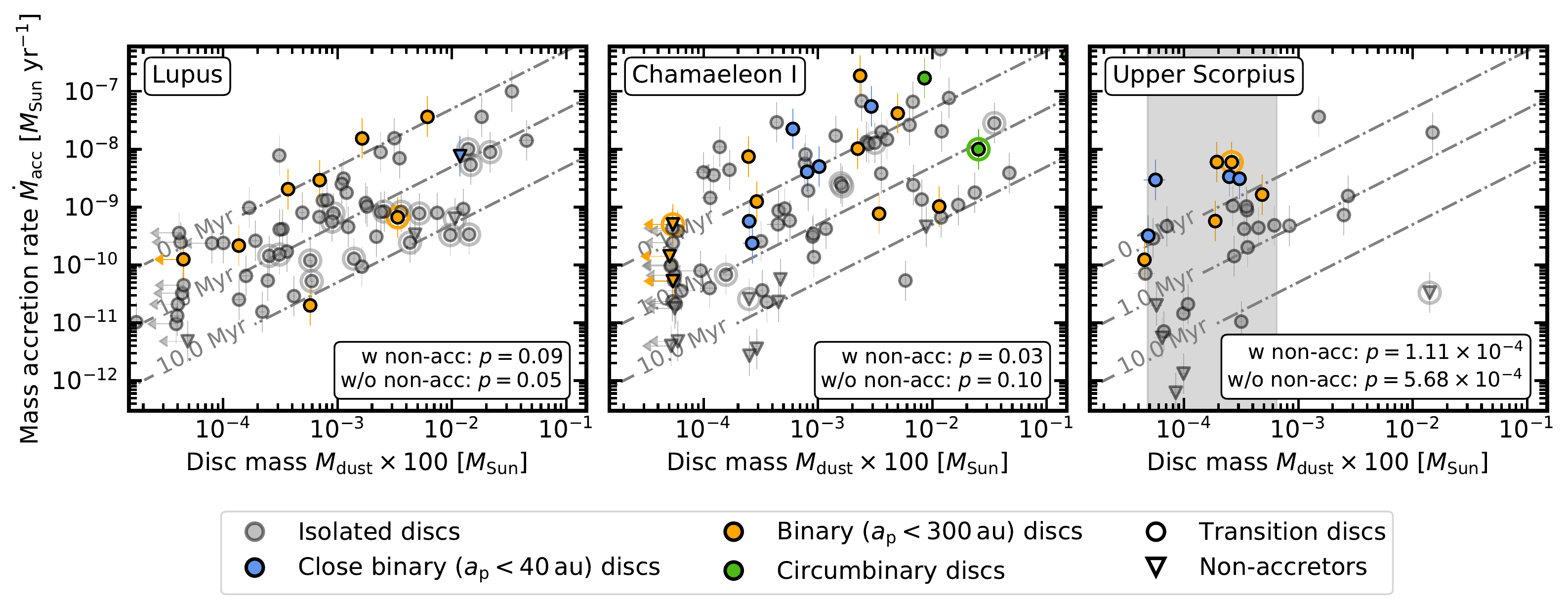}
    \caption{Same as in the left-hand panel of Fig.~\ref{fig:2.1}, with each star-forming region in a different sub-figure. \textbf{From left to right:} Lupus, Chamaeleon~I and Upper Sco. The 80 per cent completeness mass range \citep{Manara2020} is highlighted in grey in the latter region. Captions as in the right-panel of Fig.~\ref{fig:2.1}.}
    \label{fig:2.2}
\end{figure*}

Combining the targets from different regions  allows us to collect a large enough sample to make statistically significant comparisons between single-star and binary discs. This is a fair procedure as long as it does not introduce any multiplicity biases in the target selection. Although it can be argued that only Upper Sco discs with $5\times10^{-5}\lesssim M_\mathrm{disc}/M_\mathrm{Sun}\lesssim7\times10^{-4}$ have been surveyed for accretion rates with high completeness, such incompleteness does not discriminate between singles and binaries. Indeed, \citet{Barenfeld2019} showed that, in this region, discs in binaries and around singles span a similar (sub-)millimetre brightness range.

Finally, we briefly comment on the treatment of (sub-)millimetre emission upper limits in our sample, highlighting that the significant difference between the accretion time scale in binary and single-star discs is not the result of uncertainties associated with how dust mass upper limits are assigned to non-detections. We test this claim in the worst case scenario, i.e. where we assign to non-detections values which make the data-sets for singles and binaries as similar as possible. Performing a K-S test on the new cumulative curve gives a $p-$values of $2.22\times10^{-4}$.

\paragraph*{Analysis by star-forming region} While the combined data from Lupus, Chamaeleon~I and Upper Sco clearly demonstrates the difference between singles and binaries, it washes out possible evolutionary effects resulting from the different ages of the regions. For this reason, in Fig.~\ref{fig:2.2} the mass accretion rate, $\dot{M}_\mathrm{acc}$, is plotted as a function of the disc mass, $M_\mathrm{disc}=M_\mathrm{dust}\times100$ for Lupus, Chamaeleon~I and Upper Sco individually. 

Fig.~\ref{fig:2.2} shows that, while binary discs in the younger regions belong to all the previously identified families (A to C), only family A is represented in Upper Sco. To understand whether this is a disc evolution or sample-selection effect, further data are needed. We perform a K-S test on Lupus, Chamaeleon~I and Upper Sco separately, finding that the null-hypothesis that single-star and binary discs are drawn from the same $t_\mathrm{acc}$ distribution is rejected only in the latter (the K-S test $p-$values are reported in the bottom right of each panel in Fig.~\ref{fig:2.2}). Again, assessing whether this is an evolutionary or sample-selection effect needs a higher level of completeness in Upper Sco and a homogeneous analysis of the stellar multiplicity among the three star-forming regions (see Appendix~\ref{app:1}). \\

\noindent To sum up, our data show that binary discs have statistically shorter accretion time scales than those around single stars. Further data and information on disc multiplicity are needed to compare different star-forming regions from the evolutionary point of view. 

\section{Numerical methods}\label{sec:3}
In the remainder of this paper our focus is on exploring if the observed trend of a shorter accretion time scale in binaries is compatible with the expectations of models for gas and dust evolution in viscously evolving discs subject to tidal truncation by a stellar companion. To do so, we run a number of binary disc simulations that are post-processed to compute their (sub-)millimetre dust emission profiles. Throughout our analysis we use these synthetic observations to evaluate model disc properties (e.g., masses and accretion time scales) employing the same standard assumptions used in the literature (cf \citealt{Sellek2020} in the case of single-star discs). Those properties are then compared with real data. 

We model gas and dust evolution in circumstellar binary discs, neglecting re-supply of material from beyond the binary orbit and assuming that the disc cannot spread outside the tidal truncation limit imposed by the companion. For the gas we consider viscous evolution and omit the effects of photo-evaporation. For the dust the two-population model of \citet{Birnstiel2012} is employed. As a post-processing step, dust emission and accretion rates from our models are used to generate synthetic observations to be compared with real data.  

The one-dimensional finite-differences code developed by \citet{Booth2017} is used, modified to take into account tidal truncation as \citet{Zagaria2021_theo} did. Briefly, the viscous-diffusion equation \citep{Lynden-Bell&Pringle1974} is solved on a grid made up of 250 cells equally spaced in $R^{1/2}$, assuming zero-flux \citep{Bath&Pringle1981,Rosotti&Clarke2018} at the outer boundary, $R=R_\mathrm{trunc}$. We choose an exponentially-tapered power-law initial condition:
\begin{equation}\label{eq:3}
    \Sigma(R,t=0)=\dfrac{M_0}{2\pi R_0R}\exp\left(-\dfrac{R}{R_0}\right),
\end{equation}
where $R_0$ is a characteristic scale radius, and $M_0$ is the initial disc mass within $R_\mathrm{trunc}$. We explore the region of the parameter space corresponding to $R_0=10$ and $100\,\mathrm{au}$, and $M_0=1$, 3, 10, 30 and $100\,M_\mathrm{Jup}$ \citep{Sellek2020}.

A passively irradiated time-independent disc temperature profile \citep{Chiang&Goldreich1997} is considered, radially decaying as $R^{-1/2}$ and with reference temperature $T_0=88.23$~K at 10~au, calibrated on a Solar-mass star. The \citet{Shakura&Sunyaev1973} prescription is used for viscosity: we set $10^{-4}\leq\alpha\leq10^{-3}$, both for consistency with previous works \citep{Sellek2020} and because at these viscosities similar models reproduce the flux-radius correlation most closely \citep{Rosotti2019_fr,Zagaria2021_obs,Zormpas2022}.

For the dust, a uniform initial dust fraction of 0.01 is employed. The routine for grain growth implements the simplified treatment of \citet{Birnstiel2012}. In short, at each disc radius two dust populations are evolved in time: a population of small grains, whose size is the monomer grain size, $a_\mathrm{min}=0.1\,\mu\mathrm{m}$, and a population of large grains, dominating the mass, whose size, $a_\text{max}$, is determined by the combined effect of grain growth (with a grain-growth efficiency $f_\mathrm{grow}=1$, \citealt{Booth&Owen2020,Sellek2020}), fragmentation (with fragmentation velocity $u_\mathrm{f}=10\,\mathrm{m}\,\mathrm{s}^{-1}$, \citealt{Gundlach&Blum2015}) and radial drift. We refer to \citet{Birnstiel2012} for more information and to \citet{Booth2017} for the specific implementation. Finally, the dust fraction is advected along the gas flow, following the prescription of \citet{Laibe&Price2014} and taking into account the dust back-reaction on gas \citep[e.g.,][]{Dipierro2018,Garate2020}.

We compare models and observations in the data space. This means that our simulation results are post-processed to determine how they would look if observed with ALMA at the same wavelength as in \citet{Ansdell2016} and \citet{Barenfeld2016}. Their mass is then computed as in these surveys and compared with real data. We call this quantity the \textit{observer's equivalent} disc mass \citep{Sellek2020}. Doing so, we can be agnostic on the assumptions made to compute disc masses in the observations and our results would not be influenced by the latter. Our procedure is as follows. For each of our models, we first determine a synthetic surface brightness, $S_\mathrm{b}$, at frequency $\nu\approx338.75\,{\rm GHz}$ (ALMA band 7):
\begin{equation}\label{eq:4}
    S_\text{b}(R)=B_\nu(T)\bigl\{1-\exp{(-\kappa_\nu\Sigma_\text{d})}\bigr\},
\end{equation}
where $\Sigma_\text{d}$ is the dust surface density, $B_\nu$ is the black body radiation spectrum at temperature $T$ and $\kappa_\nu$ the dust (absorption) opacity, computed as in \citet{Tazzari2016} and \citet{Rosotti2019_radii}. We assumed face-on discs. If one considers the distribution of disc inclinations on the sky, it can be shown that this approximation is correct within a factor of $\langle\cos i\rangle=\pi/4\approx0.8$. Moreover, \citet{Tazzari2017} showed that correcting for the disc inclination improves the disc masses in \citet{Ansdell2016} by less than a factor of 2. Then, the \textit{observer's equivalent} dust mass, $M_\mathrm{dust}$, is computed as in Eq.~\ref{eq:2}, where the disc luminosity, $L_\nu=d^2F_\nu$, is given by the surface integral of the synthetic surface brightness. 

\section{Results}\label{sec:4}
In this Section we outline our results, discussing how the models are affected by the initial parameters. Subsequently, models and observations are confronted, in order to assess their compatibility.

\begin{figure*}
    \centering
    \includegraphics[width=\textwidth]{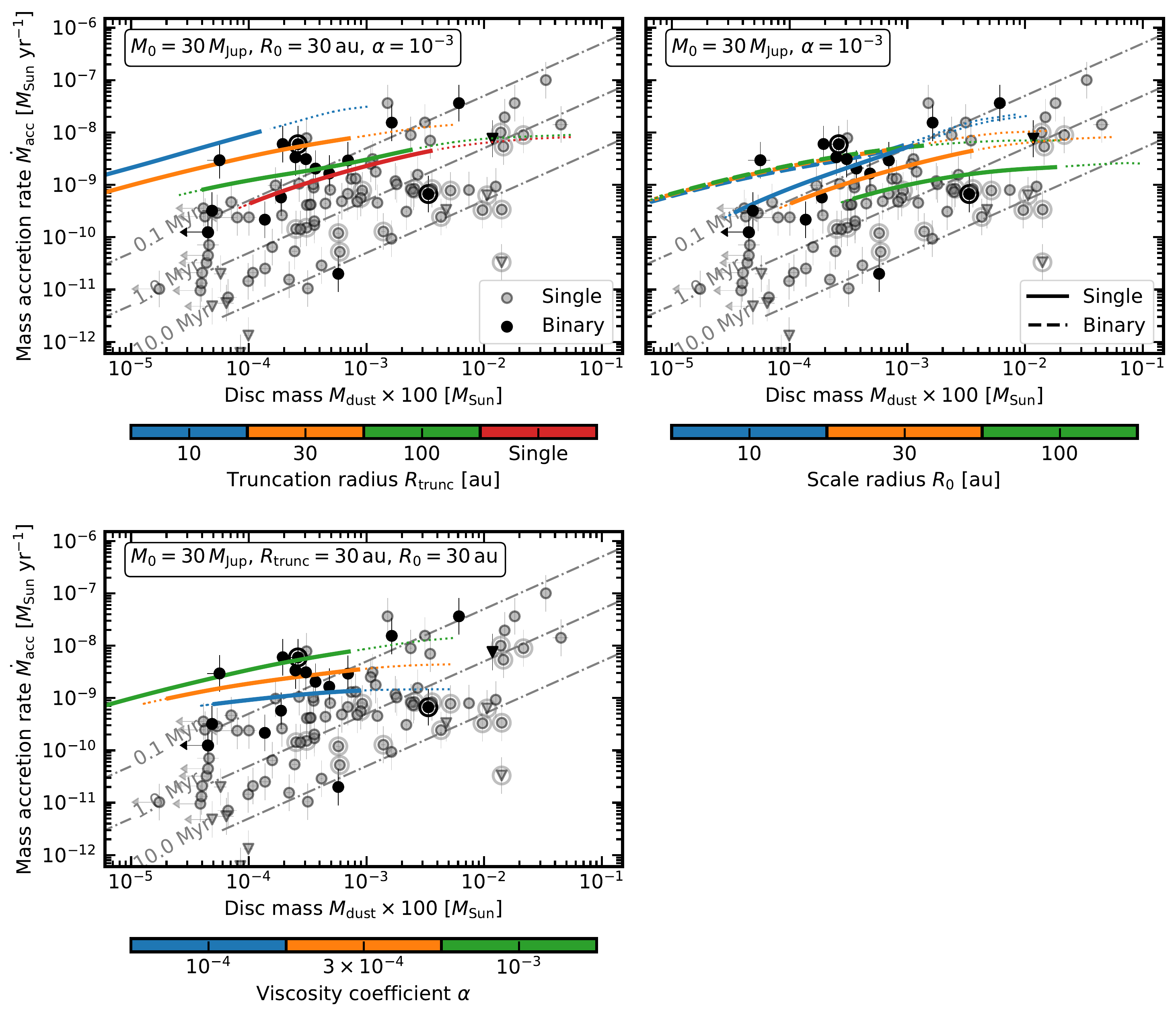}
    \caption{Mass accretion rate, $\dot{M}_\mathrm{acc}$, as a function of the observer's equivalent disc mass (see Eq.~\ref{eq:2}), $M_\mathrm{dust}\times100$. The solid tracks identify the $1\,\mathrm{Myr}\leq t\leq10\,\mathrm{Myr}$ evolutionary interval, while the dotted lines describe the evolution outside this time interval. The dependence on the truncation radius, $R_\mathrm{trunc}$ (top left), the scale radius, $R_0$ (top right), and viscosity, $\alpha$ (bottom left), are explored. In the top-right panel, solid lines refer to singles, dashed to binaries. In the background, lines of constant $t_\mathrm{acc}=0.1,\,1.0\,\mathrm{and}\,10.0\,\mathrm{Myr}$. Lupus and Upper Sco data are shown in grey for singles and black for binaries. Larger circles identify transition discs and downward-pointing triangles are used for non-accretors.}
    \label{fig:4.1}
\end{figure*}

\subsection{Model dependence on the initial disc parameters}
As a starting point, we illustrate how our results depend on the initial disc parameters. In Fig.~\ref{fig:4.1} the mass accretion rate, $\dot{M}_\mathrm{acc}$, is shown as a function of the observer's equivalent disc mass, $M_\mathrm{dust}\times 100$. The colour dotted lines identify model evolutionary tracks and their solid sections highlight the $1\,\mathrm{Myr}\leq t\leq10\,\mathrm{Myr}$ age interval. In the background, lines of constant accretion time scale for $t_\mathrm{acc}=0.1,\,1.0\,\mathrm{and}\,10\,\mathrm{Myr}$ are displayed as grey dashed-dotted lines. Lupus and Upper Sco discs are also plotted in grey for singles and black for binaries, respectively.

Let us begin with the upper-left panel of Fig.~\ref{fig:4.1}, where models with different values of the tidal truncation radius, $R_\mathrm{trunc}$, are plotted for $M_0=30\,M_\mathrm{Jup}$\footnote{We adjust the surface density normalisation so as to achieve a constant total disc mass, $M_0$, within $R_\mathrm{trunc}$. Our models are gravitationally stable, with \citet{Toomre1964} factor $Q\gtrsim1$, for $R_\mathrm{trunc}\gtrsim10\,\mathrm{au}$.}, $R_0=30\,\mathrm{au}$ and $\alpha=10^{-3}$. It can be seen from the plot that the models populate different regions of the plane. In particular, as the truncation radius decreases, the accretion time scale, $t_\mathrm{acc}$, also decreases. This reduction in $t_\mathrm{acc}$ with $R_\mathrm{trunc}$ is apparent even in the initial conditions and results from the fact that the initial accretion rate is higher for a compact, high density disc and the millimetre flux is somewhat smaller due to optical depth effects. For $t\gg0$, this trend is determined by binary discs both being fainter and accreting more. The former effect is due to the faster drift of grains for smaller values of $R_\mathrm{trunc}$ \citep{Zagaria2021_theo}, which precipitates mass depletion \citep{Zagaria2021_obs};  the latter is a consequence of the smaller viscous time scale of truncated discs \citep{Rosotti&Clarke2018}. Remarkably, while in the case of single-star models, evolutionary tracks are limited by $t_\mathrm{acc}=0.1\,\mathrm{Myr}$ for $R_0=10\,\mathrm{au}$ \citep{Sellek2020}, when $R_\mathrm{trunc}\leq100\,\mathrm{au}$, the binary models are able to populate the region of the data-plane with shorter accretion time scales, where many of the observed sources lie.

\begin{figure*}
    \centering
    \includegraphics[width=\textwidth]{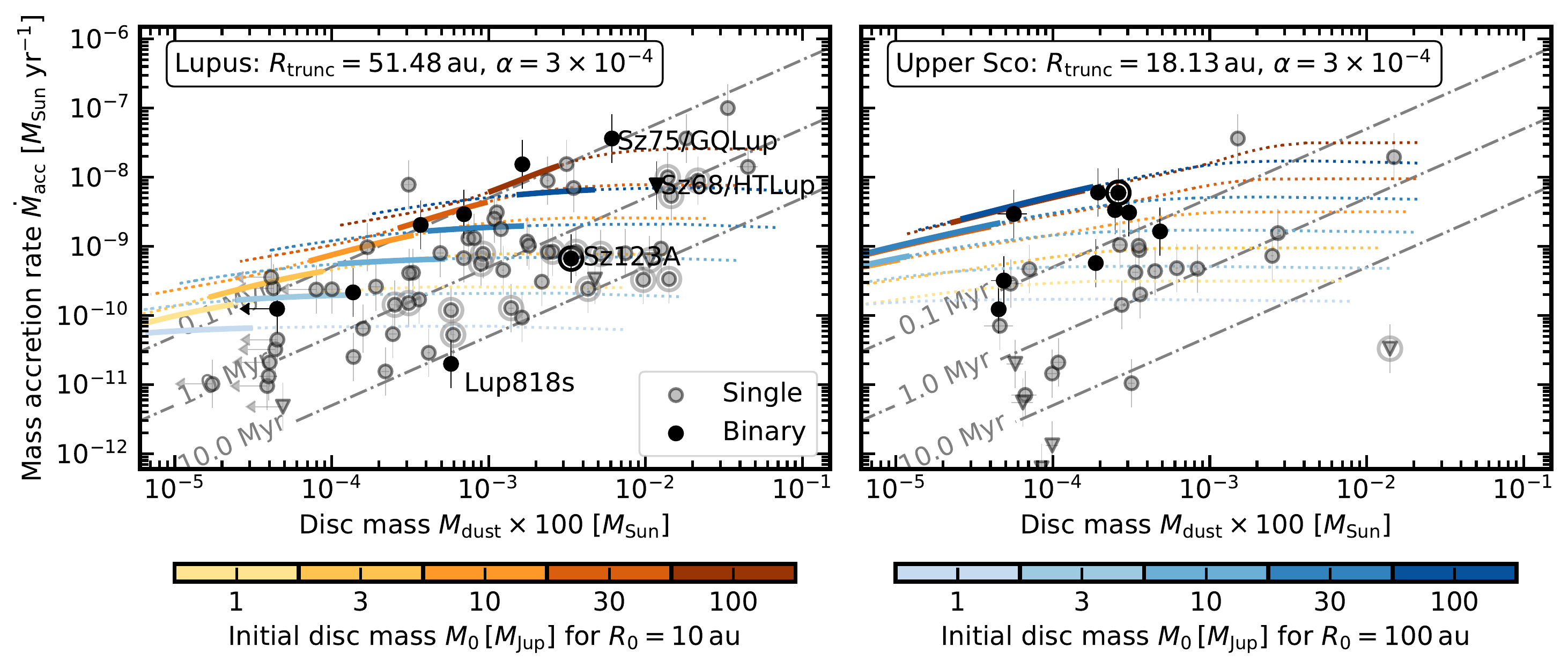}
    \caption{Mass accretion rate, $\dot{M}_\mathrm{acc}$, as a function of the observer's equivalent disc mass (see Eq.~\ref{eq:2}), $M_\mathrm{dust}\times 100$, for models with $\alpha=3\times10^{-4}$, $R_\mathrm{trunc}=51.48\,\mathrm{au}$ and $R_\mathrm{trunc}=18.13\,\mathrm{au}$ on the left and right panel, respectively. In the background, lines of constant $t_\mathrm{acc}=0.1,\,1.0\,\mathrm{and}\,10.0\,\mathrm{Myr}$. \textbf{Left:} The solid tracks identify the $1\,\mathrm{Myr}\leq t\leq3\,\mathrm{Myr}$ evolutionary interval, while the dotted lines describe the evolution outside this time interval; Lupus discs are shown in grey for singles and black for binaries. \textbf{Right:} The solid tracks identify the $5\,\mathrm{Myr}\leq t\leq10\,\mathrm{Myr}$ evolutionary interval, while the dotted lines describe the evolution outside this time interval; Upper Sco discs are shown in grey for singles and black for binaries. Larger circles identify transition discs and downward-pointing triangles are used for non-accretors. The colour bars refer to both panels.}
    \label{fig:4.2}
\end{figure*}

We remark that the short $t_\mathrm{acc}$ in the models do {\it not} imply that the gas reservoir of the disc is severely depleted but instead that the discs have evolved to low (sub-)millimetre fluxes as a result of dust radial drift. The actual gas masses in our models are up to a factor of $1000$ larger than what would be obtained by assuming a standard dust-to-gas ratio. This effect is discussed extensively in \citet{Sellek2020} and we note that it is more marked in binary discs due to the way that tidal truncation accelerates radial drift.

In the upper-right panel of Fig.~\ref{fig:4.1}, models with different values of the initial disc scale radius, $R_0$, are plotted for $M_0=30\,M_\mathrm{Jup}$ and $\alpha=10^{-3}$, in the case of a single-star disc (solid lines) and a binary disc with $R_\mathrm{trunc}=50\,\mathrm{au}$ (dashed lines). The plot shows that the tidally truncated models are only slightly influenced by the scale radius. In fact, while in single-star models $R_0$ sets the initial viscous time scale, determining the scaling of $\dot{M}_\mathrm{acc}$ \citep[e.g.,][]{Lodato2017}, and the efficiency of radial drift \citep{Birnstiel2012}, such a time scale is only determined by $R_\mathrm{trunc}$ in close binary discs \citep{Rosotti&Clarke2018,Zagaria2021_theo}. 

Finally, let us move on to the bottom, where the model dependence on viscosity, $\alpha$, is explored. The observer's equivalent dust mass is hardly affected by the viscous parameter when discs are young. However, by the age of Upper Sco, (sub-)millimetre fluxes are reduced more for higher values of $\alpha$, because the faster evolution of the gas raises the Stokes number and accelerates the radial drift of grains. This trend is observed also in single-star discs \citep{Sellek2020}, although to a lesser extent. Instead, the accretion rates depend more on viscosity and increase with $\alpha$, because it sets the velocity for gas transport thought the disc.

\subsection{Confronting models and data}
We compare binary disc models and observations in Lupus and Upper Sco. Although the Chamaeleon~I disc population and stellar multiplicity are well studied, the similar-age Lupus region is preferred because of its lower scatter in accretion time scale, better constrained (sub-)millimetre emission (85.7 \textit{vs} 79.8 per cent detections), and evenly censused stellar multiplicity (see details in Appendix~\ref{app:1}). 

A systematic comparison between models and observations requires the disc truncation radius to be fixed to some value inferred from the data. As dealing with each source individually would be too complex (because of the difficulty of estimating individual truncation radii given the unknown binary eccentricity and projection effects: see Section 5.3), our models enforce a single value of the tidal truncation radius in each region. This is determined as $R_\mathrm{trunc}=0.33\times\langle a_\mathrm{p}\rangle$, where $\langle a_\mathrm{p}\rangle$ is the average binary separation in the data, and the previous relation applies to equal mass pairs in co-planar circular orbit \citep[e.g.,][]{Paczynski1977}. As a result of a systematically closer binary population in Upper Sco, the values of $R_\mathrm{trunc}$ are different in the two regions. For this reason, we do not discuss any evolutionary dependence between Lupus and Upper Sco discs in multiple stellar systems, which would be possible only if the binary separation distributions in the two regions were very similar.

Our results are displayed in Fig.~\ref{fig:4.2} where the mass accretion rate, $\dot{M}_\mathrm{acc}$, is plotted as a function of the observer's equivalent disc mass, $M_\mathrm{dust}\times 100$, for models with $\alpha=3\times10^{-4}$, and different values of the initial disc mass, $M_0$, and scale radius, $R_0$. Lines of constant $t_\mathrm{acc}$ are displayed as dashed-dotted lines. In the left-hand panel their solid sections refer to the $1\,\mathrm{Myr}\leq t\leq3\,\mathrm{Myr}$ age range, to be compared with Lupus data, while in the right-hand panel to the $5\,\mathrm{Myr}\leq t\leq10\,\mathrm{Myr}$ age range to be compared with Upper Sco data. The dotted lines describe the evolution outside these time intervals. 

As expected, tidal truncation causes models to pass through the region of short accretion time scale occupied by the bulk of the binary data (region A in Fig.~\ref{fig:2.1}). This is due to dust being depleted more rapidly, because truncation prevents the retention of solids at large radii, which can otherwise resupply the disc at late times (see \citealt{Sellek2020_photo} for a discussion of how external photo-evaporation similarly leads to a more rapid depletion of solids by radial drift). In fact, the highest accretors in Upper Sco can be explained using viscous models with radial dust drift as long as multiplicity effects are considered. 

Nevertheless, both in Lupus and Upper Sco, model evolutionary tracks under-predict the observed disc masses. While in Lupus this issue concerns only few sources, particularly with long accretion time scale, the bulk of the Upper Sco population is roughly ten times brighter than our models. Searching the literature provides possible explanations for Lupus outliers: some of them are sub-structured (e.g., GQ~Lup and HT~Lup), while others could be affected by thermal winds (e.g., Lup~818s), as is discussed in detail in Appendix~\ref{app:2}.

In Upper Sco the under-prediction of the (sub-)millimetre fluxes by 5 to $10\,\mathrm{Myr}$ models is rather marked and applies to most of the sources in the region. Clearly, this also depends on uncertainties in opacity, although it should be noted that the assumptions we made on the grain composition correspond to an opacity at the upper end of the possible range of variation (cf. e.g., \citealt{Birnstiel2018}), and therefore with any other assumption the theoretical tracks would shift to even lower masses, exacerbating the problem. A further reason that the observed fluxes could be higher than in the models could be that (sub-)millimetre observations of closer binaries could contain emission from two discs, since more than a half of binary pairs are not resolved at the angular resolution ($\approx0.45\,\mathrm{arcsec}$) of \citet{Barenfeld2016}. This would however boost the emission by at most a factor of 2, which is not enough to account for the differences with our models. Alternatively, an evolutionary argument would suggest that only the most massive binary discs survive until the age of Upper Sco. However, models with $R_\mathrm{trunc}$ values that are too large for the observed binary separations would be required to match the data, as shown in the upper-left panel in Fig.~\ref{fig:4.1}.

To sum up, tidally-truncated models of viscously evolving discs can reproduce the short accretion time scales in the bulk of the observed binary population. Halting the fast drift of solids is required to account for the (sub-)millimetre brightness of Upper Sco and some sources in Lupus.

\section{Discussion}\label{sec:5}

\subsection{Sub-structures in binary discs}
The first detection of gaps and rings in HL~Tau \citep{ALMA_HLTau} provided striking evidence of sub-structures in planet-forming discs (see \citealt{Andrews2020} for a review). Despite such structures often being assumed to be ubiquitous \citep{Andrews2018_DSHARP,Long2018}, their presence/absence in compact or low mass discs is yet to be established
(e.g., \citealt{Long2019}, Jennings et al. subm.).

Sub-structures have been detected in binary discs as well: e.g., spirals in HT~Lup~A and AS~205~N \citep{Kurtovic2018}, inner cavities in XZ~Tau~B \citep{Osorio2016}, UZ~Tau~E and CIDA~9A \citep{Long2019,Manara2019}, gaps and rings in AS~205~S \citep{Kurtovic2018}, GQ~Lup~A \citep{Long2020} and T~Tau~N \citep{Yamaguchi2021}. However, it is not clear if they are as common as in isolated discs. In fact, since binary discs are systematically smaller than those in single stars \citep{Manara2019}, identifying sub-structures in the former is more challenging than in the latter.

Here we focus on axisymmetric annular features and discuss their influence on the correlation between mass accretion rates and disc masses. In the case of single-star discs, \citet{Sellek2020} suggested that bright rings, consistent with trapping large grains in local pressure maxima \citep{Pinilla2012,Dullemond2018}, could be an explanation for discs with long accretion time scales too massive for their smooth models (region C in Fig.~\ref{fig:2.1}). It is then tempting to hypothesise that binary discs in region C are likewise sub-structured and their longer accretion time scale compared with those in region A could be similarly explained. However, while for single-star discs this inference is supported by the presence of imaged gaps and rings, none of our binary sources in region C has clearly detected gaps or cavities, making such systems preferential targets for a first higher-resolution analysis of sub-structures in binary discs.

Noticeably, whereas discs around single stars fill the region between population B and C continuously, disc in binaries appear to be well separated: for accretion rates $\lesssim10^{-10} M_\mathrm{Sun}\,\mathrm{yr}^{-1}$ there is a lack of such discs over about an order of magnitude in disc mass (see Fig.~\ref{fig:2.1}). While it cannot be ruled out that this effect is driven by the smaller sample size of discs in binaries, it may hint at a dichotomy of evolutionary paths. Binary discs containing sub-structures may remain over several Myr in region C, comparable to the situation in structured single-star discs. Instead, in the absence of sub-structures, our models predict a more rapid decay of disc luminosity in binaries than in singles (see top-left panel of Fig.~\ref{fig:4.1}), particularly for discs with low initial disc masses \citep{Birnstiel2012}. The presence of dust traps could thus explain the apparent lack of binary discs with low accretion rates and intermediate disc masses.

Such a picture, both for binaries and singles, implies that traps maintain discs in region C over time scales of a few Myr since sources are found in region C in both Lupus and Chamaeleon I which are approximately this age. Notably, this region is almost devoid of sources in Upper Sco. This would suggest that sources do not evolve \textit{vertically} downwards from region C, since otherwise Upper Sco would contain several systems with high (sub-)millimetre fluxes but low/undetected accretion rates. Alternatively dust traps may be disrupted on a time scale of 5 to $10\,\mathrm{Myr}$ \citep{Rosotti2013,Sellek2020}, so that systems evolve out of region C \textit{horizontally} to the left. Evidence of remnant sub-structures in Upper Sco discs would support such an interpretation. Furthermore, our understanding of the late time trajectory of discs in the plane of accretion rates vs (sub-)millimetre fluxes would be improved by increasing the completeness of accretion rate determinations in Upper Sco for the discs with the highest (sub-)millimetre fluxes.

The most intriguing interpretation of dust gaps is that of being carved by planets in the act of formation \citep[e.g.,][]{Zhang2018}. If we retain the hypothesis that region C sources (whether binary or single) are sub-structured, then it is tempting to extend to binaries the argument of \citet{Manara2019_pps} that discs in region C are forming giant planets, whose deep gaps not only halt radial drift but also reduce accretion. The existence of some binaries in region C would thus support the idea that giant planet formation may be under way in at least some of these discs.

We also consider whether sub-structures can explain the fact that the observed binaries in Upper Sco are too bright compared with model predictions (see right-hand panel of Fig.~\ref{fig:4.2}, contrasting the data, shown as black dots, with the models, displayed as solid curves). If this were the case then we would need dust traps to be almost ubiquitous. However, apart from GQ~Lup~A \citep{Long2020}, a source $\approx10$ times more massive than those in Upper Sco, we have no evidence of imaged gaps and rings in any of our binary discs. This can be easily explained by the limited angular resolution of the relevant surveys\footnote{$0.25\,\mathrm{arcsec}$ \citep{Ansdell2018}, $0.60\,\mathrm{arcsec}$ \citep{Pascucci2016} and $0.45\,\mathrm{arcsec}$ \citep{Barenfeld2016}; to be compared with, e.g., the $0.12\,\mathrm{arcsec}$ resolution of \citet{Long2019}.}.

A more compelling problem is that in this picture sub-structured discs would live in region A, very far from region C where they would be expected according to our previous argument. However, looking at single-star sources in region C highlights that all transition discs with imaged cavities are clustered here, suggesting that these sub-structures play a role in maintaining conditions not only of high disc masses but also relatively low accretion rates. We can then hypothesise that, in the binary as in the single-star case, while discs in region C are characterised by prominent, large scale sub-structures affecting both gas and dust evolution, shallower, small scale dust traps would be required in the case of Upper Sco, halting dust drift but not impacting substantially the gas.

Testing this hypothesis quantitatively requires detailed modelling of secular binary evolution in the presence of sub-structures, as well as higher-resolution observations in Upper Sco.

\subsection{Coagulation efficiency in binary discs}
Shallow dust traps at small radial locations are not the only  solution to the mass budget problem in Upper Sco. This can be alternatively explained in terms of a reduced grain growth efficiency in close binary discs. 

Tidal interactions between gravitationally bound stars and their discs can be at the origin of asymmetries, spiral arms, eccentric modes, vertical hydraulic jumps, mass transfer between discs and relative shocks... (e.g., \citealt{Nelson2000,Muller&Kley2012,Picogna&Marzari2013}). It is commonly hypothesised that these higher-dimensional effects could halt grain growth in circumstellar discs orbiting close binaries ($a\leq30\text{ to }50\,\mathrm{au}$, \citealt{Nelson2000,Picogna&Marzari2013})\footnote{Given the average separation of Lupus and Upper Sco binaries, this argument applies preferentially to discs in the latter region.}. When spirals are triggered by tidal interactions, the disc temperature rises enough for water and other volatile species to be vaporised. As ices make up to 60 per cent of our grain composition \citep{Tazzari2016,Rosotti2019_radii}, vaporisation would lead to solid dis-aggregation and thus promote fragmentation\footnote{Adopting $u_\mathrm{f}=10\,\mathrm{m}\,\mathrm{s}^{-1}$ as in Section~\ref{sec:3} is only appropriate for ice-coated grains \citep{Gundlach&Blum2015}, while $u_\mathrm{f}=1\,\mathrm{m}\,\mathrm{s}^{-1}$ is more suited for silicate-rich grains \citep{Blum&Wurm2008}.}. Even though grain growth could take place between spirals, it would then need to begin from material in gaseous phase and would be disrupted by the frequent interactions with the spirals themselves (expected to occur on a time scale of less than $100\,\mathrm{yr}$, \citealt{Nelson2000}). 

\begin{figure}
    \centering
    \includegraphics[width=\columnwidth]{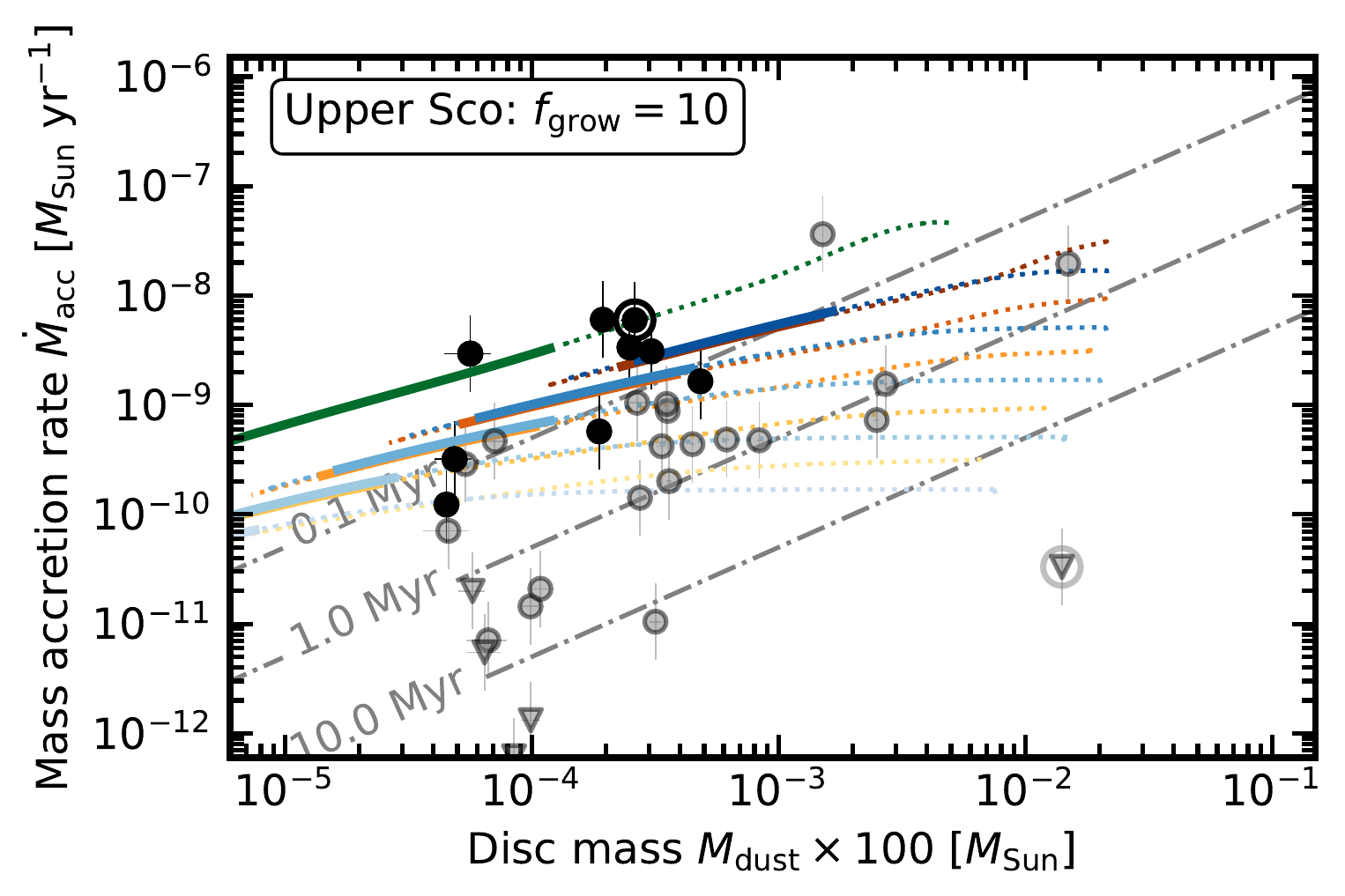}
    \caption{Same as in the right panel of Fig.~\ref{fig:4.2}, assuming $f_\mathrm{grow}=10$. The green line identifies an additional model
    ($R_0=30\,\mathrm{au}$ and $M_0=100\,M_\mathrm{Jup}$) with a smaller truncation radius corresponding to the closest binary in Upper Sco: $R_\mathrm{trunc}=6.7\,\mathrm{au}$.}
    \label{fig:5.1}
\end{figure}

\begin{figure*}
    \centering
    \includegraphics[width=\textwidth]{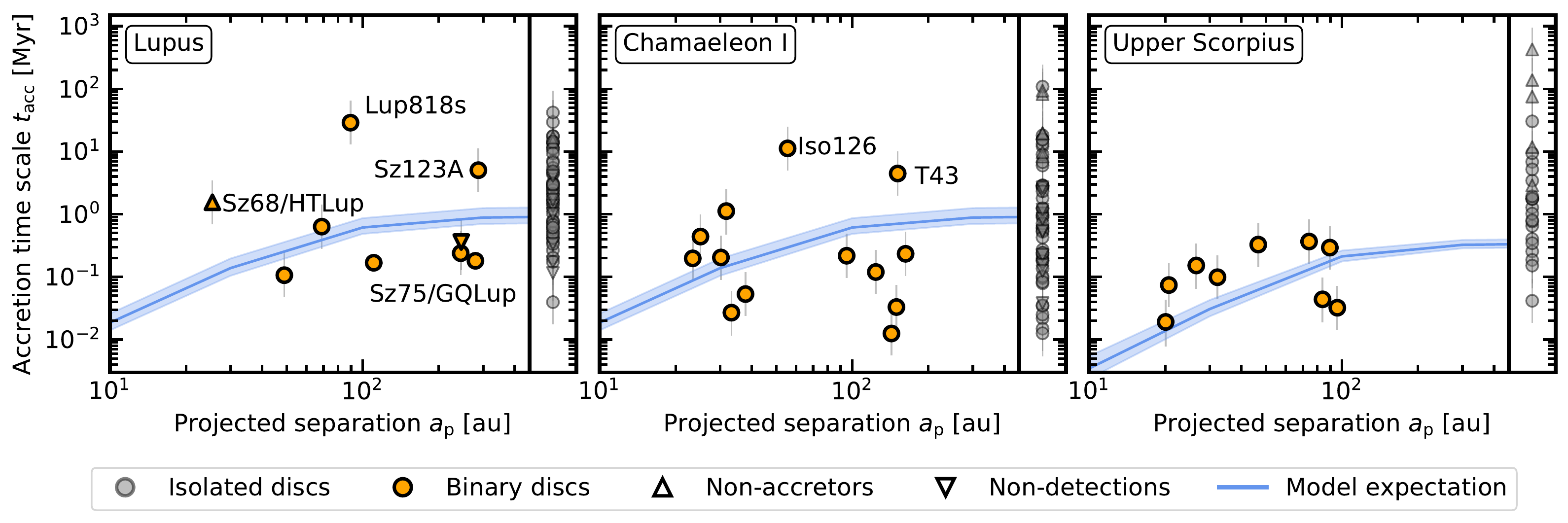}
    \caption{Accretion time scale, $t_\mathrm{acc}$, as a function of the pair projected separation, $a_\mathrm{p}$, in Lupus, Chamaeleon~I and Upper Sco. Orange dots are used for discs in binary systems and grey dots for discs evolving in isolation. The blue tracks highlight our models expectations for $M_0=30\,M_\mathrm{Jup},\,R_0=30\,\mathrm{au},\,\alpha=3\times10^{-4}$.}
    \label{fig:5.2}
\end{figure*}

These effects can be taken into account in our one-dimensional models by reducing the grain growth efficiency. To do so, we introduce a coagulation efficiency factor, $f_\mathrm{grow}$ \citep{Booth&Owen2020,Sellek2020}. So far we assumed $f_\mathrm{grow}=1$, meaning that all grain collisions lead to coagulation. Instead, if $f_\mathrm{grow}>1$, only a fraction $f_\mathrm{grow}^{-1}$ of the collisions result in growth, and sticking is thus less efficient, as expected in binary discs. A lower coagulation efficiency would also increase the drift time scale, given the balance between dust growth and drift achieved in the drift-dominated regime \citep{Birnstiel2012}. Recently, \citet{Booth&Owen2020} suggested that assuming $f_\mathrm{grow}=10$ could be beneficial for explaining the depletion of refractory elements in the Sun relative to nearby Solar twins. \citet{Sellek2020} ran single-star disc models with $f_\mathrm{grow}=10$, showing that they were able to reproduce massive Lupus discs that they could not explain assuming more efficient growth. However $f_\mathrm{grow}=10$ single star models could not match the data in Upper Sco, being too massive at any given accretion rate.

We retain $f_\mathrm{grow}=10$ as in these works because it provides a balance between fast growth, where solids drift fast, and slow growth, where grains are fragmentation-dominated throughout and the dust follows the gas \citep{Booth&Owen2020}. The results of our binary disc models with $f_\mathrm{grow}=10$, in the case of $R_\mathrm{trunc}=18.13\,\mathrm{au}$ and $\alpha=3\times10^{-3}$, are shown in Fig.~\ref{fig:5.1}. Regardless of the initial disc parameters, our models evolve towards disc masses that are ten times higher than for $f_\mathrm{grow}=1$ (as seen by \citealt{Sellek2020}). This difference reconciles binary models and observations in Upper Sco, with the caveat that the highest accretors are still not reproduced. This is easily explained by our choice of an average binary separation. Enforcing $R_\mathrm{trunc}=6.7\,\mathrm{au}$, corresponding to the minimum binary separation in Upper Sco, allows  models to reproduce high accretors as well. This is shown in the representative case of $R_0=30\,\mathrm{au}$ and $M_0=100\,M_\mathrm{Jup}$, depicted with a solid green line in Fig.~\ref{fig:5.1}.

\subsection{Accretion time scale \textit{vs} binary projected separation}
Using one-dimensional viscous models, we have shown that the fact that proto-planetary discs in multiple systems occupy region A in Fig.~\ref{fig:2.1} (where $t_\mathrm{acc}\approx0.1\,\mathrm{Myr}$) can be reproduced by the secular effects of outer disc truncation on dust and gas evolution. Such models suggest that a shorter accretion time scale in discs in closer binaries should be expected, owing to the larger accretion rates and lower dust masses predicted in those systems (see e.g., top-left panel of Fig.~\ref{fig:4.1}).

Let us therefore examine how the observationally-inferred accretion time scale, $t_\mathrm{acc}$, depends on the measured projected binary separation, $a_\mathrm{p}$. This is displayed in Fig.~\ref{fig:5.2} for Lupus, Chamaeleon~I and Upper Sco binary discs, using orange dots. Single-star discs are also shown for comparison, using grey dots. No particular trend of $t_\mathrm{acc}$ with $a_\mathrm{p}$ can be recognised in the data, at odds with our expectations. These are highlighted by the blue tracks and shaded area for the representative case with $M_0=30\,M_\mathrm{Jup},\,R_0=30\,\mathrm{au},\,\alpha=3\times10^{-4}$ and its $1\sigma$ spread.

However, what primarily sets the accretion time scale in our models is the location where truncation occurs. In fact, this is not determined by the binary projected separation only, but is expected to be affected by other effects, such as the relative inclination of the interacting pair with respect to the plane of the sky \citep{Manara2019}, the binary mass ratio and eccentricity \citep[e.g.,][]{Paczynski1977,Artymowicz&Lubow1994,Pichardo2005}, and the relative misalignment between the disc and the binary plane \citep{Lubow2015}.

In particular, comparing the binary separation distribution in our data with the observed relation between main sequence binaries eccentricity and orbital period \citep{Raghavan2010}, would imply that some discs in our sample could have a significant eccentricity, thus  washing out the expected dependence of $t_\mathrm{acc}$ on $a_\mathrm{p}$. Recent evidence that eccentricity in wider binaries follows a thermal rather than uniform distribution (see \citealt{Hwang2021}, Fig.~6) would suggest that the most eccentric pairs would be those with larger $a_\mathrm{p}$. As long as the bulk of the population has relatively smaller eccentricities ($e\approx0.3$ as expected in the field, e.g., \citealt{Duchene&Kraus2013}), this is not in tension with the fact that the the dust disc sizes in Taurus and $\rho$~Ophiuchus can be well explained assuming low eccentric orbits (\citealt{Zagaria2021_obs} and \citealt{Rota2022}). 

Even though the comparison between models and specific sources can be challenged by uncertainties on binary eccentricities, we do not expect them to substantially modify our results in Section~\ref{sec:4}, because they would reduce the average truncation radius only by a factor of a few and this is potentially compensated by the underestimation of the binary separation by its projection on the plane of the sky. Nevertheless, recovering the theoretically-expected trend of $t_\mathrm{acc}$ with $a_\mathrm{p}$ is only part of the problem, given the large scatter in Fig.~\ref{fig:5.2}. In fact, on top of the previously mentioned mechanisms, other effects related to grain coagulation efficiency, disc evolution and dispersal can influence the accretion time scale. We do not explore further these possibilities because, given our uncertainty on the determination of $R_\mathrm{trunc}$, modelling would be under constrained by the data. Finally, some of the sources in Fig.~\ref{fig:5.2} can also be components of undetected higher-order multiples.

\section{Summary and conclusions}\label{sec:6}
In this paper we made use of gas and dust one-dimensional simulations to address how stellar multiplicity can influence the correlation between mass accretion rates onto forming stars and the masses of their proto-planetary discs. 

Hereafter our main conclusions are summarised:
\begin{itemize}
    \item We searched the literature for information on the stellar multiplicity of Lupus, Chamaeleon~I and Upper Sco discs with both (sub-)millimetre emission and accretion rates in \citet{Manara2016,Manara2020} and \citet{Mulders2017}. We found that roughly 20 per cent of the targeted discs are in binary or higher-order multiple stellar systems;
    \item We introduced the accretion time scale, $t_{\rm acc}$, as the ratio of observationally estimated disc mass to accretion rate (see Eq.~\ref{eq:1}) and found that discs in gravitationally bound pairs have a systematically shorter accretion time scale than discs evolving in isolation, clustering around $t_\mathrm{acc}\approx0.1\,\mathrm{Myr}$. The short accretion time scales is a consequence both of the well known fact that dust masses are lower in binaries compared with single stars \citep[e.g.,][]{Harris2012} {\it and} of the fact that accretion rates are somewhat higher than in single stars (see Fig.~\ref{fig:2.1});
    \item To understand the trend in the data, we ran models of tidally truncated proto-planetary discs subject to viscous evolution, grain growth and radial drift. We found that our models show shorter accretion time scales for lower values of $R_\mathrm{trunc}$, as a consequence of the faster radial drift and reduced disc lifetime determined by tidal truncation (see upper-left panel of Fig.~\ref{fig:4.1});
    \item Models and data agree reasonably well in Lupus (see left-hand panel of Fig.~\ref{fig:4.2}); we suggest that a small number of anomalously (sub-)millimetre bright discs may be explained by the existence of sub-structures, while a disc with anomalously long $t_\mathrm{acc}$, may potentially undergo internal photo-evaporation;
    \item In Upper Sco our models systematically under-predict disc masses by up to a factor of ten (see right-hand panel of Fig.~\ref{fig:4.2}). It is unclear if this could be the effect of undetected traps halting the drift of solids. We also find that the agreement is improved if the grain growth efficiency is reduced (see Fig.~\ref{fig:5.1}) as has been hypothesised in the dynamically active environments of discs in binaries:
    \item Finally, we find no clear evidence that the observationally-inferred accretion time scale increases with the binary projected separation (see Fig.~\ref{fig:5.2}), at odds with our theoretical expectation that the accretion time scale should increase with the tidal truncation radius of the disc. However, the relationship between projected separation and tidal truncation radius is complicated by projection effects and the orbital eccentricity: given the relatively limited dynamic range of projected separations in the sample, this lack of correlation is therefore not necessarily surprising.
\end{itemize}

\section*{Acknowledgements}
We thank the anonymous reviewer for their comments that helped to improve the manuscript. FZ acknowledges support from STFC and Cambridge Trust for a Ph.D. studentship. GR acknowledges support from the Netherlands Organisation for Scientific Research (NWO, program number
016.Veni.192.233) and from an STFC Ernest Rutherford Fellowship
(grant number ST/T003855/1). This work was partly supported by the Deutsche Forschungs-Gemeinschaft (DFG, German Research Foundation) - Ref no. FOR 2634/1 TE 1024/1-1. This work has also been supported by the European Union’s Horizon 2020 research and innovation programme under the Marie Sklodowska Curie grant agreement number 823823 (DUSTBUSTERS).

Software: \texttt{numpy} \citep{numpy20_2020Natur.585..357H}, \texttt{matplotlib} \citep{matplotlib_Hunter:2007}, \texttt{scipy} \citep{scipy_2020SciPy-NMeth}, \texttt{JupyterNotebook} \citep{Jupyter_nootbok}. This research has made use of the SIMBAD database, operated at CDS, Strasbourg, France \citep{Wenger2000}, and the Binary Star Database by Kendra Kellogg, Lowell Observatory. Colours in Fig.s~\ref{fig:4.2} and \ref{fig:5.1} are from \texttt{ColorBrewer.org} by Cynthia A. Brewer, Geography, Pennsylvania State University.


\section*{Data Availability}
The code used in this paper is publicly available on GitHub at
\texttt{github.com/rbooth200/DiscEvolution}. The data underlying
this paper are available in the ALMA archive as explained in the papers quoted in Section~\ref{sec:2} and Appendix~\ref{app:1}.


\bibliographystyle{mnras}
\bibliography{refs}


\appendix

\section{Sample selection -- multiplicity}\label{app:1}
Discs in multiple stellar systems are identified as follows.

In Lupus we rely on the near-infrared, adaptive optics $L'$ band VLT/NACO survey of \citet[see \citealt{Zurlo2020} for details on the observation technique and data reduction]{Zurlo2021}. Their sample includes all the pre-main-sequence stars observed with ALMA by \citet{Ansdell2016,Ansdell2018}. Companions were identified using the \texttt{daophot} algorithm in the \texttt{photutils} package in the \texttt{astropy} library \citep{astropy} and then inspected visually. Stellar pairs with separation less than $20\,\mathrm{au}$ (roughly $0.1\,\mathrm{arcsec}$ at Lupus distance) could not be resolved. Of the sources not in the sample of \citet{Ansdell2016}, Sz~75/GQ~Lup and Sz~77 are close binaries with separation of $0.7\,\mathrm{arcsec}$ \citep{Neuhauser2005}, and $1.8\,\mathrm{arcsec}$ \citep{Ghez1997}, respectively. 11 discs in our sample are in multiple systems closer than 2~arcsec. 

In Chamaeleon~I, not all the sources taken into account by \citet{Pascucci2016} have been surveyed, to our knowledge. \citet{Lafreniere2008} looked for companions to the previously identified Chamaeleon~I members brighter than $K=13$ of \citet{Luhman2004}, using $K$ and $H$ band VLT/NACO observations with median $K$ band PSF FWHM of $0.086\,\mathrm{arcsec}$. Several of these targets were re-observed by \citet{Daemgen2013} with a similar $K$ band magnitude selection criterion and angular resolution, confirming their multiplicity. DI~Cha/2MASS~J11072074-7738073 and Sz~22/2MASS~J11075792-7738449 were re-imaged with NACO in the $J$, $H$ and $K$ bands at three and two different epochs, allowing for the identification of closer companions, as explained by \citet{Schmidt2013}. 2MASS~J11175211-7629392 is also a binary with a separation of $2\,\mathrm{arcsec}$ \citep{Manara2017}. Using VLTI/PIONIER at $H$ band with a maximum angular resolution of $3\,\mathrm{mas}$, \citet{Anthonioz2015} identified WW~Cha/2MASS~J11100010-7634578 as a close binary system with circumbinary disc. Similarly, \citet{Nguyen2012}, using high-resolution RV measurements suggested that FM~Cha/2MASS~J11095340-7634255 and CS~Cha/2MASS~J11022491-7733357 are spectroscopic binaries with separation less than $0.8\,\mathrm{au}$ and $4.0\,\mathrm{au}$ \citep{Guenther2007}, respectively\footnote{TW~Cha/2MASS~J10590108-7722407 was also identified as a candidate spectroscopic binary by \citet{Nguyen2012}, but to our knowledge this source was not further investigated in the next years. The cavity in the CS~Cha disc is reasonably due to its binary nature \citep[e.g.,][]{Ginski2018}, hence it is not considered a TD.}. 21 discs in our sample are in multiple systems closer than $2\,\mathrm{arcsec}$. 

In Upper Sco we relied on \citet{Barenfeld2019} AO imaging and non-redundant aperture masking $K$ band observations obtained using the NIRC2 AO imager on the Keck~II telescope. Their sample includes all the pre-main-sequence stars observed with ALMA by \citet{Barenfeld2016}. Following \citet{Kraus2008}, any sources brighter than $K=15$ and within $2\,\mathrm{arcsec}$ of a target star are considered to be candidate bound companions. 9 discs in our sample are in multiple systems closer than $2\,\mathrm{arcsec}$. 

In our analysis we consider as binary systems all those gravitationally bound pairs whose projected separation is closer than $2\,\mathrm{arcsec}$ at the average distance of Upper Sco, that is to say $a_\mathrm{p}\approx300\,\mathrm{au}$. This corresponds to approximately $1.83\,\mathrm{arcsec}$ at the average distance of Lupus and $1.53\,\mathrm{arcsec}$ at the average distance of Chamaeleon~I. 9 discs in Lupus (excluding Sz~123~B because sub-luminous), 16 in Chamaeleon~I (excluding the 3 known circumbinary discs) and 9 in Upper Sco are in those systems and have measured disc masses and accretion rates. We also define close binaries as pairs with projected separation less than $40\,\mathrm{au}$, following \citet{Kraus2012}. Table~\ref{tab:A1} summarises our literature search outcomes for binaries in Lupus, Chamaeleon~I and Upper Sco, and lists the average projected separation\footnote{In the case of multiple systems each disc-bearing binary pair (see the definition of \citealt{Harris2012}) was considered.} and inferred truncation radius to be used in models.

\begin{table}
    \centering
    \begin{tabular}{|c|c|c|c|c|c|}
    \hline
     & Discs & \multicolumn{2}{c}{Binaries} & $\langle a_\mathrm{p}\rangle$ & $R_\mathrm{trunc}$ \\
     & & $<300\,\mathrm{au}$ & $<40\,\mathrm{au}$ & (arcsec) & (au) \\
    \hline
    Lupus        & 70 & 9 & 1 & 0.98 & 51.84 \\[2.5pt]
    Chamaeleon~I & 84 & 16 & 6 & 0.43 & 27.18 \\[2.5pt]
    Upper Sco    & 34 & 9  & 4 & 0.42 & 18.13 \\
    \hline
    \end{tabular}
    \caption{Binary statistics, average projected separation and truncation radius in Lupus, Chamaeleon~I and Upper Sco.}
    \label{tab:A1}
\end{table}

\paragraph*{Single-star discs are really singles?} Despite Lupus, Chamaeleon~I and Upper Sco stars having been extensively searched for companions, their stellar multiplicity was not homogeneously determined. This makes assessing whether differences between binary discs in these star-forming regions are intrinsic or determined by selection biases harder. 

For example, we would expect fewer close (accreting) binaries (with discs) in more evolved regions \citep[e.g.,][]{Kraus2012}. However, while in the Lupus $9/70\approx12.8\%$ discs are in multiples, this percentage roughly doubles to $9/34\approx26.5\%$ in the under-complete Upper Sco sample. This picture worsens if pairs closer than $40\,\mathrm{au}$ \citep{Kraus2012} are considered: $1/70\approx1.4\%$ in Lupus and $4/34\approx11.8\%$ in Upper Sco. Both binary samples were primarily compiled using AO imaging \citep{Zurlo2021,Barenfeld2019}. However, in Upper Sco \citet{Barenfeld2019} made also use of non-redundant aperture masking, able to achieve higher contrast in the case of very close ($\approx100\,\mathrm{mas}$) binaries.

A homogeneous study of disc multiplicity between star-forming regions of different ages is needed to properly assess the role of stellar companions to affect disc evolution and to compare theoretical expectations in regions of different ages.

\section{Lupus outliers}\label{app:2}
Hereafter we comment in detail on the discs in Lupus that cannot be reproduced by our smooth binary models.

In Sz~75/GQ~Lup a broad inner gap ($\approx10\,\mathrm{au}$) and a tentative outer gap ($\approx32\,\mathrm{au}$) were imaged by \citet{Long2020}: dust trapping can be a motivation for its mass being higher than expected. Lup~818s has an unusually long accretion time scale ($t_\mathrm{acc}\approx50\,\mathrm{Myr}$). This could be a result of photo-evaporation which, if included, would cause accretion rates to plummet at late times \citep{Sellek2020}. The system parameters for Lup~818s ($a_\mathrm{p}=90\,\mathrm{au}$, \citealt{Zurlo2021}, and $M_*\approx0.057\,M_\mathrm{Sun}$, Manara et al. subm.) are such that it is expected that X-ray photo-evaporation would generate an inner hole and cut off accretion (see \citealt{Rosotti&Clarke2018}, Fig.~3). According to the results of \citet{Rosotti&Clarke2018}, X-ray photo-evaporation could also be important in Sz~123~A, a well known SED transition disc with azimuthal asymmetries \citep{Ansdell2018}, and $\approx 1.77\,\mathrm{arcsec}$-close companion \citep{Zurlo2021}. Sz~68/HT~Lup~A is the closest, most massive binary disc in Lupus. Its behaviour could be explained by its age, $\log(t\,\mathrm{yr}^{-1})=5.9\pm0.3$ \citep{Andrews2018_DSHARP}, being lower than the population-average one. However, it is also possible that this disc was born massive. It shows spiral structures \citep{Kurtovic2018}, evidence that our one-dimensional models are probably not suited to reproduce its behaviour.

\bsp	
\label{lastpage}
\end{document}